\documentclass[prl,twocolumn,floatfix,a4paper,amssymb,showpacs,superscriptaddress,groupedaddress]{revtex4-2}  
\usepackage[english]{babel}
\usepackage[T1]{fontenc}
\usepackage{inputenc}
\usepackage{amsmath}
\usepackage{amsfonts}
\usepackage{amssymb}
\usepackage{graphicx}
\usepackage{dsfont}
\usepackage{tikz}
\usepackage[percent]{overpic}
\usepackage{xcolor}
\usepackage{physics}
\usepackage[caption = false]{subfig}
\usepackage{babelbib}
\usepackage{wrapfig}
\usepackage{pgfplots}
\usepackage{babelbib}
\usepackage{hyperref}
\usepackage{lipsum}
\usepackage{transparent}
\usepackage{amsthm}
\usepackage{listings}
\usepackage{eso-pic}
\usepackage{pgfplots}
\usepackage{dcolumn}   
\usepackage{bm}  
\usepackage{natbib}

\begin{document}
	\title{Efficient Time-Domain Approach for Linear Response Functions}	
	\author{Michel Panhans}
	\affiliation{Center for Advancing Electronics Dresden, Technische Universit\"at Dresden, 01062 Dresden, Germany}
	\affiliation{Department of Chemistry, Technische Universit\"at M\"unchen, 85748 Garching b. M\"unchen, Germany}
	\author{Frank Ortmann}
	\affiliation{Center for Advancing Electronics Dresden, Technische Universit\"at Dresden, 01062 Dresden, Germany}
	\affiliation{Department of Chemistry, Technische Universit\"at M\"unchen, 85748 Garching b. M\"unchen, Germany}
	
	\date{\today}
	
	\begin{abstract}
   We derive the general Kubo formula in a form that solely utilizes the time evolution of displacement operators. The derivation is based on the decomposition of the linear response function into its time-symmetric and time-antisymmetric part. We relate this form to the well-known fluctuation-dissipation formula and discuss theoretical and numerical aspects of it. The approach is illustrated with an analytical example for magnetic resonance as well as a numerical example where we analyze the electrical conductivity tensor and the Chern insulating state of the disordered Haldane model. We introduce a highly efficient time-domain approach that describes the quantum dynamics of the resistivity of this model with an at least 1000-fold better performance in comparison to existing time-evolution schemes.
	\end{abstract}
	\maketitle

\paragraph{Introduction.}
The Kubo formalism \cite{Kubo1957} is a powerful and universal theoretical tool to connect the complex microscopic dynamics of condensed matter systems with their macroscopic thermodynamic properties. Within this framework, linear susceptibilities relate any physical observable to any other perturbing forces exerted by the experimenter and can thus explain diverse material properties such as electrical conductivity or magnetic susceptibility. Many discoveries such as the quantum Hall effects \cite{BASTIN19711811,Streda_1982,AOKI1981,Yu61}, intrinsic spin Hall effects \cite{MurakamiPRL2004,Sinova2004} or quantum spin Hall effects  \cite{KaneMele,QiPRB2006,Bernevig1757} have been explained using linear response theory. In addition, it is used to describe magnetic resonance absorption \cite{KuboTomita1954}, the theory of the anomalous Hall effect in the Dirac equation \cite{CrepieuxPRB2001}, the bulk viscosity of quark-gluon matter \cite{KARSCH2008217}, the thermal conductivity of disordered harmonic solids \cite{Sevin_li_2019} or linear absorption spectra in metals and semiconductors \cite{bechstedt2016many}.\\
The Kubo formalism allows for large-scale numerical calculations to describe quantum systems, which elude an analytical description such as complex nanoscale systems \cite{van2013scaling,markussen2006electronic,latil2005electronic} or when disorder or electronic correlations are present. Great efforts were therefore spent in the last two decades to work on efficient numerical implementations of Kubo formulae in the field of electronic transport \cite{WeisseRMP2006,FAN2020}. Many numerical implementations of Kubo formulae \cite{Roche1997,Fratini2011,Ishii2011,LandsteinerPRL2011,PlumaricPRC2012,Ishii2014,OrtmannPRB2015,Garcia2015PRL} have been developed to optimize the description and better understand the transport physics of quantum systems and connect it to macroscopic transport phenomena and experiments. A key requirement of efficient numerical strategies is to avoid the diagonalization of the Hamiltonian matrix and elaborate on linear scaling approaches, which is particularly challenging in case of off-diagonal tensor components of the response function \cite{FAN2020}.\\
In this paper, we present a decomposition of linear response functions into a time-symmetric and a time-antisymmetric part and find that they can solely be expressed in terms of displacement operators at equal time, which eventually enables an efficient implementation and computation. We relate this representation to the well-known form of the quantum fluctuation-dissipation theorem \cite{Kubo1966} that directly connects susceptibilities and power spectra. We find a natural generalization of the description of cross-power spectra of arbitrary pairs of perturbation forces and response observables. This enables the efficient description of transverse response phenomena such as Hall effects, spin Hall effects, or other tensor quantities (e.g.~transversal magnetization effects, anisotropic diffusion-tensors, etc.) at the same footing as the longitudinal response. This unified description makes the development of specific algorithms unnecessary. 

\paragraph{Kubo formulae.}
The general Kubo formula for the linear response of observable $A$ in presence of a small but time-dependent perturbation $H'(t)$  of the quantum system with the Hamiltonian $\hat{H}=\hat{H}_0+\hat{H}'(t)$ can be written as \cite{Kubo1957}
\begin{align}
\text{Tr}\left(\hat{\rho}(t)\hat{A}\right)=\text{Tr}\left(\hat{\rho}_0\hat{A}\right)+\frac{i}{\hbar}\int\limits_{-\infty}^tdt'\,\text{Tr}\left(\hat{\rho}_0\left[\hat{H}_{\text{I}}'(t'),\hat{A}(t)\right]\right) \label{general Kubo}
\end{align}
with the time-dependent density operator $\hat{\rho}(t)$ that is driven by the unperturbed Hamiltonian $\hat{H}_0$ and the equilibrium density operator $\hat{\rho}_0$ (canonical or grand canonical \footnote{{For the canonical ensemble, we use $\hat{\rho}_0=e^{-\beta\hat{H}}/\text{Tr}(e^{-\beta\hat{H}})$ and for the grand-canonical ensemble we use $\hat{\rho}_0=e^{-\beta\left(\hat{H}-\mu\hat{N}\right)}/\text{Tr}(e^{-\beta\left(\hat{H}-\mu\hat{N}\right)})$. }}) that describes the quantum system  in  absence  of  the  perturbation. $\hat{H}_{\text{I}}'(t')=e^{it'\hat{H}_0/\hbar}\hat{H}'(t')e^{-it'/\hat{H}_0/\hbar}$ is the conventional perturbation operator in the interaction picture and $\hat{A}(t)=e^{it\hat{H}_0/\hbar}\hat{A}(0)e^{-it\hat{H}_0/\hbar}$ 
is the Heisenberg time evolution of $\hat{A}$.\\
If the perturbation is characterized by an arbitrary time-dependent modulation function $F(t)$ coupled to an operator $\hat{B}(0)$, i.e.~$\hat{H}'(t)=-F(t)\hat{B}(0)$, the general Kubo formula reads
\begin{align}
\begin{aligned}
\text{Tr}\left(\hat{\rho}(t)\hat{A}\right)
&=\text{Tr}\left(\hat{\rho}_0\hat{A}\right)+\int\limits^{\infty}_0dt'\,F(t-t')f_{AB}(t')
 \label{general Kubo 2}\,.
\end{aligned}
\end{align}
with the response function 
\begin{align}
\begin{aligned}
f_{AB}(t)&=-\frac{i}{\hbar}\text{Tr}\left(\hat{\rho}_0\left[\hat{B}(0),\hat{A}(t)\right]\right).
\label{response function}
\end{aligned}
\end{align}
While a great number of examples for perturbations of this form exist, such as electric or magnetic fields, we emphasize that the following results do not assume a special form of the Hermitian operators $\hat{A}$ and $\hat{B}$.
Central objects in our study are the displacement operators $\Delta\hat{A}(t)$ and $\Delta\hat{B}(t)$ that are defined as $\Delta\hat{A}(t)=\hat{A}(t)-\hat{A}(0)$ and $\Delta\hat{B}(t)=\hat{B}(t)-\hat{B}(0)$\footnote{The term displacement operator appears in different contexts in physics with different meanings. For clarity, with displacement operator, we mean the difference of two operators: one is the Heisenberg-evolved operator and the other is its non-time-evolved operator. The general use of this term here is not restricted to any specific operator.}. The first result is the following theorem.\\

$\bf{Theorem\, 1:}$ The response function $f_{AB}(t)$ can be written in the form
\begin{align}
\begin{aligned}
f_{AB}(t)-f_{AB}(0)&=\frac{1}{2\hbar}\mathcal{D}^-_{AB}(t)+\frac{1}{2\hbar}\tan\left(\frac{\beta\hbar}{2}\frac{d}{dt}\right)\mathcal{D}^+_{AB}(t)\\
& \label{Generalized Kubo displacement form}
\end{aligned}
\end{align}
where we have defined the displacement operator anticommutator function (DAF)
\begin{align}
\mathcal{D}^+_{AB}(t)&=\text{Tr}\left(\hat{\rho}_0\left\{\Delta\hat{A}(t),\Delta\hat{B}(t)\right\}\right) \label{DAF0}
\end{align}
and the displacement operator commutator function (DCF)
\begin{align}
\mathcal{D}^-_{AB}(t)&=-i\,\text{Tr}\left(\hat{\rho}_0\left[\Delta\hat{A}(t),\Delta\hat{B}(t)\right]\right) \,.\label{CDF0}
\end{align}
in which the square brackets and the curly brackets denote the commutator and the anticommutator, respectively. \\
One frequently encounters a special situation, namely that the response function of interest reads $f_{\dot{A}B}(t)$, i.e.~it includes an observable that is associated to a time derivative $\dot{A}$. In such cases we obtain a related theorem.

$\bf{Theorem \,2:}$ Consider that, if additionally to the assumptions of Theorem 1, the response function of interest $f_{\dot{A}B}(t)$ contains an observable defined as  $\dot{\hat{A}}=\left(i/\hbar\right)\left[\hat{H}_0,\hat{A}\right]$, the linear response $\text{Tr}\left(\hat{\rho}_0(t)\dot{\hat{A}}\right)$ can be written as
\begin{align}
 \text{Tr}\left(\hat{\rho}_0(t)\dot{\hat{A}}\right)=\int\limits_0^{\infty}dt\,F(t-t')f_{\dot{A}B}(t')
\end{align}
with
\begin{align}
\begin{aligned}
f_{\dot{A}B}(t)=\frac{1}{2\hbar}\frac{d}{dt}\mathcal{D}_{\dot{A}B}^-(t)+\frac{1}{2\hbar}\tan\left(\frac{\beta\hbar}{2}\frac{d}{dt}\right)\frac{d}{dt}\mathcal{D}_{\dot{A}B}^+(t)
\,. \label{Generalized Einstein}
\end{aligned}
\end{align}
The proof of the theorems is provided in the appendix.\\ 

Equation \eqref{Generalized Kubo displacement form} is a specific representation of the linear response function $f_{AB} (t)$ of an arbitrary pair of observables $A$ and $B$,  whose connection to the conventional cross-correlation function will be shown further below. Owing to symmetry relations $\mathcal{D}_{AB}^- (t)=\mathcal{D}_{AB}^- (-t)$ and $\mathcal{D}_{AB}^+ (t)=\mathcal{D}_{AB}^+ (-t)$, the response function $f_{AB} (t)$ is expressed by a decomposition into a time-symmetric part $f_{AB}^{\text{ts}}(t)=f_{AB}(0)+\mathcal{D}_{AB}^- (t)/2\hbar$ and a time-anti-symmetric part $f_{AB}^{\text{ta}} (t)=\tan\left(\beta\hbar/2\, d/dt\right)\mathcal{D}_{AB}^+(t)/2\hbar$. We note that in the special case of $\hat{B}=\hat{A}$ the response function reads $f_{AA} (t)=\tan\left(\beta\hbar/2\, d/dt\right)\mathcal{D}_{AA}^+(t)/2\hbar$; while if $\hat{B}=\dot{\hat{A}}$ one finds $f_{A\dot{A}}(t)=f_{A\dot{A}}(0)+\mathcal{D}_{A\dot{A}}^- (t)/2\hbar$. Their time evolution    therefore depend only on either of the two functions DAF or DCF in contrast to $f_{AB} (t)$ and $f_{\dot{A}B} (t)$ where both are required. 
We further emphasize that these result do not exploit any time-symmetry properties of the operators or observables $A$ and $B$ that are sometimes used to demonstrate Onsager-Casimir relations (OCR) \cite{Onsager1931,CasimirRMP1945}, but are independently obtained and valid even in absence of any time symmetry for the operators. Indeed the OCR connects the time reversal to the exchange of the operators ($\hat{A}$ and $\hat{B}$), a connection which will be discussed further below.\\
Considering the exchange of the operators in Eqs.~\eqref{DAF0} and \eqref{CDF0}, we trivially obtain $\mathcal{D}_{AB}^+ (t)=\mathcal{D}_{BA}^+(t)$ and $\mathcal{D}_{AB}^- (t)=-\mathcal{D}_{BA}^-(t)$ as well as $f_{AB}^{\text{ta}} (t)=f_{BA}^{\text{ta}}(t)$ and $f_{AB}^{\text{ts}} (t)=-f_{BA}^{\text{ts}}(t)$. Again this is different to the OCR since in the latter case a symmetry of the Hamiltonian needs to be assumed (e.g.~the magnetic field needs to be reversed), which is not the case in the above relations. The absence of this assumption allowed us to derive this more general approach.

Furthermore, $\mathcal{D}_{AB}^+(t)$ and $\mathcal{D}_{AB}^-(t)$ satisfy the Cauchy-Schwarz inequality $\left|\mathcal{D}_{AB}^{+(-)}(t)\right|\leq\sqrt{\mathcal{D}_{AA}^+(t)\mathcal{D}_{BB}^+(t)}$ as a strict upper limit for arbitrary $\hat{A}$ and $\hat{B}$. In the special case when $\hat{B}=\hat{A}$ the equality for $\mathcal{D}_{AA}^+(t)$ holds. However for $\hat{B}=\dot{\hat{A}}$ the DCF satisfies the uncertainty relation $\left|\mathcal{D}_{A\dot{A}}^{-}(t)\right|\leq\sqrt{\mathcal{D}_{AA}^+(t)\mathcal{D}_{\dot{A}\dot{A}}^+(t)}$.

\paragraph{Connection to the cross-correlation function and its time symmetry.}
The quantum version of the cross-correlation function $S_{AB}(t)$ is defined by the symmetrized cross-correlation function \cite{Kubo1966}
\begin{align}
\mathcal{S}_{AB}(t)=\frac{1}{2}\text{Tr}\left(\hat{\rho}_0\left\{\hat{B}(0),\hat{A}(t)\right\}\right),
\end{align}
which is a real valued function in accordance with the classical correlation function $\mathcal{S}_{AB}^{\text{class}}(t)=\left<\rho_0B(0)A(t)\right>$ in which $B(0)$ and $A(t)$ always commute. 
We now show that the decomposition of $\mathcal{S}_{AB}(t)$ into its time-symmetric and time-antisymmetric part results in a representation with displacements operators $\Delta\hat{A}(t)$ and $\Delta\hat{B}(t)$. \\

$\bf{Theorem \,3:}$
Using the definitions of the DAF and DCF, the decomposition of $\mathcal{S}_{AB}(t)$ into its time-symmetric and time-antisymmetric part reads
\begin{align}
\mathcal{S}_{AB}^{\text{ts}}(t)&=\mathcal{S}_{AB}(0)-\frac{1}{4}\mathcal{D}^+_{AB}(t)\label{Correlation time-symmetric},\\
\tan\left(\frac{\beta\hbar}{2}\frac{d}{dt}\right)\mathcal{S}_{AB}^{\text{ta}}(t)&=-\frac{\hbar}{2}f_{AB}(0)-\frac{1}{4}\mathcal{D}^-_{AB}(t)\label{Correlation time anti-symmetric}\,.
\end{align}
The proof of equations \eqref{Correlation time-symmetric} and \eqref{Correlation time anti-symmetric} is provided in the appendix.
The displacement-operator form for the response function in equation \eqref{Generalized Kubo displacement form}  and for the correlation function  according to equation \eqref{Correlation time-symmetric} and \eqref{Correlation time anti-symmetric} result in
\begin{align}
f_{AB}(t)=-\frac{2}{\hbar}\tan\left(\frac{\beta\hbar}{2}\frac{d}{dt}\right)\mathcal{S}_{AB}(t)\,.\label{fluctuation-dissipation theorem time domain}
\end{align} 
which is the fluctuation-dissipation theorem expressed in the time domain since it relates the response function $f_{AB}(t)$ with the cross-correlation function $\mathcal{S}_{AB}(t)$.

\begin{figure}[h!]
	\centering
	\begin{center}
		\includegraphics[width=1\linewidth]{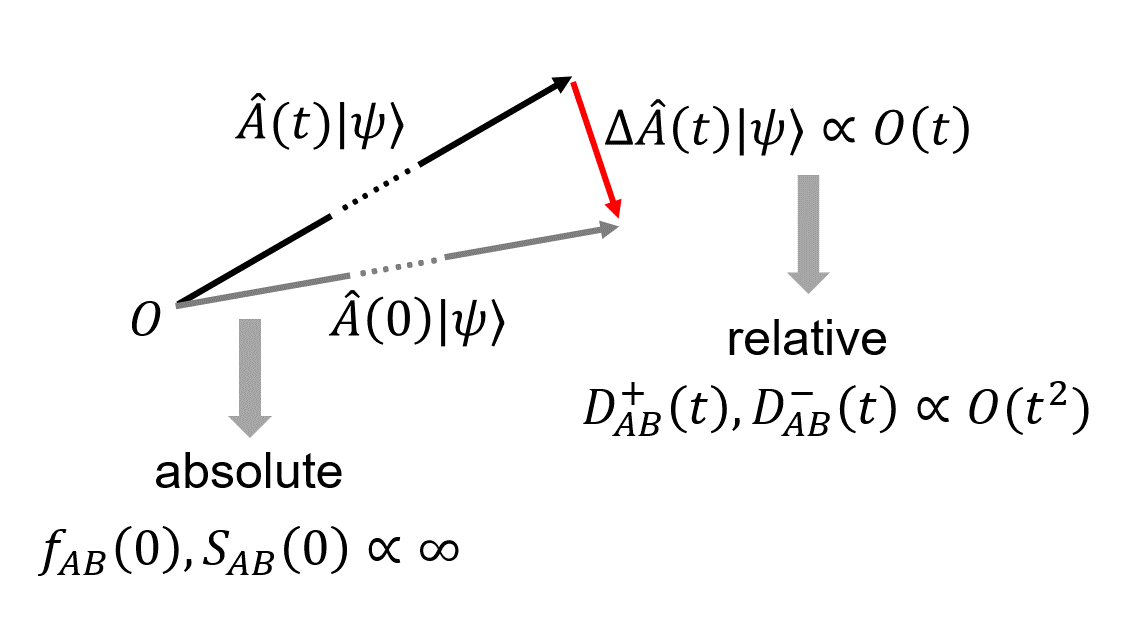}
		\caption{Geometric interpretation of the time evolution of the change of the response function  $f_{AB}(t)-f_{AB}(0)$. Only the ``small displacement vectors'' after operator action with $\Delta\hat{A}(t)$ (and $\Delta\hat{B}(t)$ equivalently) needs to be calculated instead of the ``large vectors'' related to the operators $\hat{A}(t)$ or $\hat{A}(0)$.}
		\label{fig:T-symmetry-decomposition}
	\end{center}
\end{figure}

\paragraph{Numerical aspects of possible implementations.}
Eq.~\eqref{Generalized Kubo displacement form} shows that the response function is, apart from its initial value $f_{AB}(0)$, solely determined by the simultaneous displacements that are expressed by $\Delta\hat{A}(t)$ and $\Delta\hat{B}(t)$ and not by the operators $\hat{A}(t)$ and $\hat{B}(t)$ themselves. This form has two advantages: it can be exploited efficiently in numerical approaches and it avoids possibly ill-defined quantities such as diverging expectation values. An important example for the latter is the dipole operator in periodic systems \cite{MartinPhysRevB1974}.
In figure \ref{fig:T-symmetry-decomposition} we illustrate this aspect by analogy to geometric vectors and represent the action of the operators $\hat{A}(t)$, $\hat{A}(0)$ and $\Delta\hat{A}(t)$ on an arbitrary state $\ket{\psi}$ in the Hilbert space as such vectors (with their norms $N_{\hat{A}(t)}=\left\lVert\hat{A}(t)\ket{\psi}\right\rVert$ etc.).
Instead of dealing with the ``large vectors'', i.e.~large numbers for $N_{\hat{A}(t)}$ and $N_{\hat{A}(0)}$, it is sufficient to use the ``small vector displacements'' (with small norm $N_{\Delta\hat{A}(t)}$) and avoid the calculation of quantities that are potentially extremely large (or infinite), which is always unpractical in the quantitative analysis of any macroscopic system. More precisely, we see that the DAF and the DCF are numerically well-conditioned, because they evolve as $\order{t^2}$ at the starting point $t=0$, in contrast to correlation functions that involve $\hat{A}(t)$ directly. 
Thus, we conclude that the evolution of the response function $f_{AB}(t)$ away from its reference $f_{AB}(0)$ can be expressed by only the simultaneous displacement operators of the observables $A$ and $B$, suggesting an accurate iterative time-propagation approach. 

Another aspect is of equal importance because Eq.~\eqref{Generalized Kubo displacement form} or \eqref{Generalized Einstein} allows for a {\it simultaneous} simulation of both diagonal and off-diagonal tensor components for arbitrary linear responses. This time-domain approach represents a generalization over present charge-transport approaches \cite{FAN2020}. In particular for the electrical conductivity, the time-domain approach derived in this paper complements the Kubo-Bastin formula in the energy domain, which has been successfully used to study electron transport in topological systems and twisted bilayer graphene \cite{Garcia2015PRL,AndejelkovicPRM2018}.

At the same time, the simulation of diagonal and off-diagonal matrix elements on equal footing in the same algorithm allows minimizing discrepancies due to different numerical approaches. This is particularly relevant when combining matrix elements such as for calculation of the resistivity from diagonal and off-diagonal conductivities.

\paragraph{Series expansion of the displacement functions.}
We now provide some useful relations involving the displacement functions $\mathcal{D}_{AB}^+(t)$ and  $\mathcal{D}_{AB}^-(t)$ that let us find the coefficients of their Taylor series expansion around $t = 0$.
Firstly, we find an important equivalence between the displacement functions $\mathcal{D}_{AB}^+(t)$, $\mathcal{D}_{AB}^-(t)$, $\mathcal{D}_{\dot{A}\dot{B}}^+(t)$ and $\mathcal{D}_{\dot{A}\dot{B}}^-(t)$, namely
\begin{align}
\frac{d^2}{dt^2}\mathcal{D}^+_{AB}(t)&=4\mathcal{S}_{\dot{A}\dot{B}}(0)-\mathcal{D}^+_{\dot{A}\dot{B}}(t)\label{recurrence DAF}\,,\\
\frac{d^2}{dt^2}\mathcal{D}^-_{AB}(t)&=-2\hbar f_{\dot{A}\dot{B}}(0)-\mathcal{D}^-_{\dot{A}\dot{B}}(t)\label{recurrence DCF}\,.
\end{align}
Since $\mathcal{D}_{AB}^+(t)$ and  $\mathcal{D}_{AB}^-(t)$ are both time symmetric, their Taylor series expansion contains only even powers of $t$. As a consequence, Eqs.~\eqref{recurrence DAF} and \eqref{recurrence DCF} provide a recurrence relation for their expansion coefficients. Because $\mathcal{D}_{AB}^+(t)$, $\mathcal{D}_{AB}^-(t)$ and also all higher order displacement functions, e.g.~$\mathcal{D}^+_{\ddot{A}\ddot{B}}(t)$ and $\mathcal{D}^-_{\ddot{A}\ddot{B}}(t)$ vanish at $t = 0$, we have
\begin{align}
\frac{d^{2k}}{dt^{2k}}\mathcal{D}^+_{AB}(t)\Biggr|_{t=0}&=4(-1)^{k-1}\mathcal{S}_{A^{(k)}B^{(k)}}(0)\\
\frac{d^{2k}}{dt^{2k}}\mathcal{D}^-_{AB}(t)\Biggr|_{t=0}&=2\hbar(-1)^kf_{A^{(k)}B^{(k)}}(0)
\end{align}
where $A^{(k)}$ denotes the $k$-fold nested commutator of $\hat{H}_0$ with $\hat{A}$, i.e.~$A^{(k)}=\left(i/\hbar\right)^k\left[\hat{H}_0,\hat{A}\right]_k$.
Thus, the series expansion of the DAF and the DCF can be written as
\begin{align}
\begin{aligned}
\mathcal{D}^+_{AB}(t)&=4\sum\limits_{k=1}^\infty\frac{(-1)^{k-1}t^{2k}}{(2k)!}\mathcal{S}_{A^{(k)}B^{(k)}}(0)\\
&=-4\sum\limits_{k=1}^\infty\frac{t^{2k}}{(2k)!}\mathcal{S}_{AB^{(2k)}}(0)\label{series expansion DAF}\,,
\end{aligned}\\
\begin{aligned}
\mathcal{D}^-_{AB}(t)&=2\hbar\sum\limits_{k=1}^\infty\frac{(-1)^kt^{2k}}{(2k)!}f_{A^{(k)}B^{(k)}}(0)\\
&=2\hbar\sum\limits_{k=1}^\infty\frac{t^{2k}}{(2k)!}f_{AB^{(2k)}}(0)\,.\label{series expansion DCF}
\end{aligned}
\end{align}
We show two versions for each of these intriguing expansions to highlight that different coefficients can be used to express $\mathcal{D}_{AB}^{+(-)}(t)$. They can be calculated using both the $k$-fold nested commutators of $\hat{A}$ with $\hat{H}_0$ and $\hat{B}$ with $\hat{H}_0$ or, alternatively, only the $2k$-fold  nested commutators of $\hat{B}$ with $\hat{H}_0$ (or $2k$-fold  nested commutators of $\hat{A}$ with $\hat{H}_0$). This great flexibility allows to choose the easiest way of calculation of such commutators.

\paragraph{Analytical example: Magnetic resonance of an isolated electron spin.}
As a minimal-model illustration of Theorem 1 that demonstrates how the displacement functions $\mathcal{D}_{AB}^+(t)$ and $\mathcal{D}_{AB}^-(t)$ determine the linear response, we use the spin precession under a perturbative magnetic field. The unperturbed system Hamiltonian of a single electron spin with mass $m_\text{e}$  in presence of a large magnetic field $B_z$ is described by $\hat{H}_0=\hat{\sigma}_z\mu_{\text{B}}B_z$ with the Bohr magneton $\mu_{\text{B}}=\hbar e/2m_{\text{e}}$. Here no spin-relaxation processes due to spin-spin interaction or spin-orbit interaction are considered. The system is perturbed by an adiabatically switched oscillating and linearly polarized magnetic field $\hat{H}'(t)=-\hat{\sigma}_x\mu_{\text{B}}B_xF(t'-t)$ with $B_x\ll B_z$. Its response in terms of the macroscopic expectation value of the spin vector $\bf{S}$ with the components ${S_{\alpha}}=\text{Tr}\left(\hat{\rho}\hbar\hat{\sigma}_{\alpha}/2\right)$ is monitored.\\
Within this very simple model, we can analyze analytically the individual contributions to the response function $f_{AB}(t)$ according to equations \eqref{series expansion DAF} and \eqref{series expansion DCF}. With $\hat{A}=\hbar\hat{\sigma}_{\alpha}/2$ and $\hat{B}=\hat{\sigma}_x\mu_{\text{B}}B_x$ we find
\begin{align}
f_{AB}(0)&=\mu_{\text{B}}B_x\text{Tr}\left(\hat{\rho}_0\hat{\sigma}_z\right)\delta_{\alpha y}\,,\\
\mathcal{D}^-_{AB}(t)&=2\hbar \mu_{\text{B}}B_x\text{Tr}\left(\hat{\rho}_0\hat{\sigma}_z\right)\delta_{\alpha y}\left(\cos\left(\omega_{\text{L}}t\right)-1\right)\,,\\
\mathcal{D}^+_{AB}(t)&=2\hbar\mu_{\text{B}}B_x\left(1-\cos\left(\omega_{\text{L}}t\right)\right)\text{Tr}\left(\hat{\rho}_0\hat{\sigma}^2_x\right)\delta_{\alpha x}
\end{align}
with the Larmor frequency $\omega_{\text{L}}=eB_z/m_{\text{e}}$.
The application of the operator $\tan\left(\beta\hbar/2\,d/dt\right)$ on $\mathcal{D}^+_{AB}(t)$ is evaluated analytically and for a single spin we can use $\text{Tr}\left(\hat{\rho}_0\hat{\sigma}_z\right)=\tanh\left(\beta\mu_{\text{B}}B_z\right)$ and $\text{Tr}\left(\hat{\rho}_0\hat{\sigma}_x^2\right)=1$. 
If the perturbation field oscillates with $F(t'-t)=\lim\limits_{\eta\rightarrow 0^+}e^{-i\omega t-\eta t}$ then the susceptibility tensor $\chi_{\alpha\beta}(\omega)=2\mu_{\text{B}}\mu_0S_{\alpha}(\omega)/B_{\beta}V\hbar$ has the components 
\begin{align}
\chi_{xx}(\omega)&=\frac{\mu_{\text{B}}^2\mu_0}{\hbar V}\tanh\left(\beta\mu_{\text{B}}B_z\right)\frac{2\omega_{\text{L}}}{\omega^2-\omega_{\text{L}}^2}\,,\\
\chi_{yx}(\omega)&=-\frac{\mu_{\text{B}}^2\mu_0}{\hbar V}\tanh\left(\beta\mu_{\text{B}}B_z\right)\frac{2i\omega}{\omega^2-\omega_{\text{L}}^2}\,.
\end{align}
in full consistence with the textbook expression for the paramagnetic susceptibility in the theory of magnetic resonance (MR) \cite{pottier2009nonequilibrium}. This shows that MR is correctly described with the analytical displacements functions, while the numerical evaluation allows to describe much more complex situations of the magnetic resonance and a broader class of spin-related phenomena in the new approach.
Indeed, if the system Hamiltonian $\hat{H}_0$ becomes more complex (e.g.~with additional spin-orbit interaction, i.e.~$\hat{H}_0\rightarrow\hat{H}_0+\gamma \hat{\bf{L}}\cdot\hat{\bf{S}}$, or spin-spin interactions such as in the Heisenberg model for ferro- or anti-ferromagnetic materials, i.e.~$\hat{H}_0\rightarrow\hat{H}_0+\gamma\sum_{\left<i,j\right>}\hat{{\bf{S}}}_i\cdot\hat{{\bf{S}}}_j$) the displacement functions  $\mathcal{D}^+_{AB}(t)$ and $\mathcal{D}^-_{AB}(t)$ can rarely be expanded analytically but Eq.~\eqref{Generalized Kubo displacement form} can be evaluated efficiently by time evolution approaches \cite{WeisseRMP2006,FAN2020}. In this work we showcase another application field that is charge carrier transport.

\begin{figure}[h!]
\centering
\includegraphics[width=1.05\linewidth]{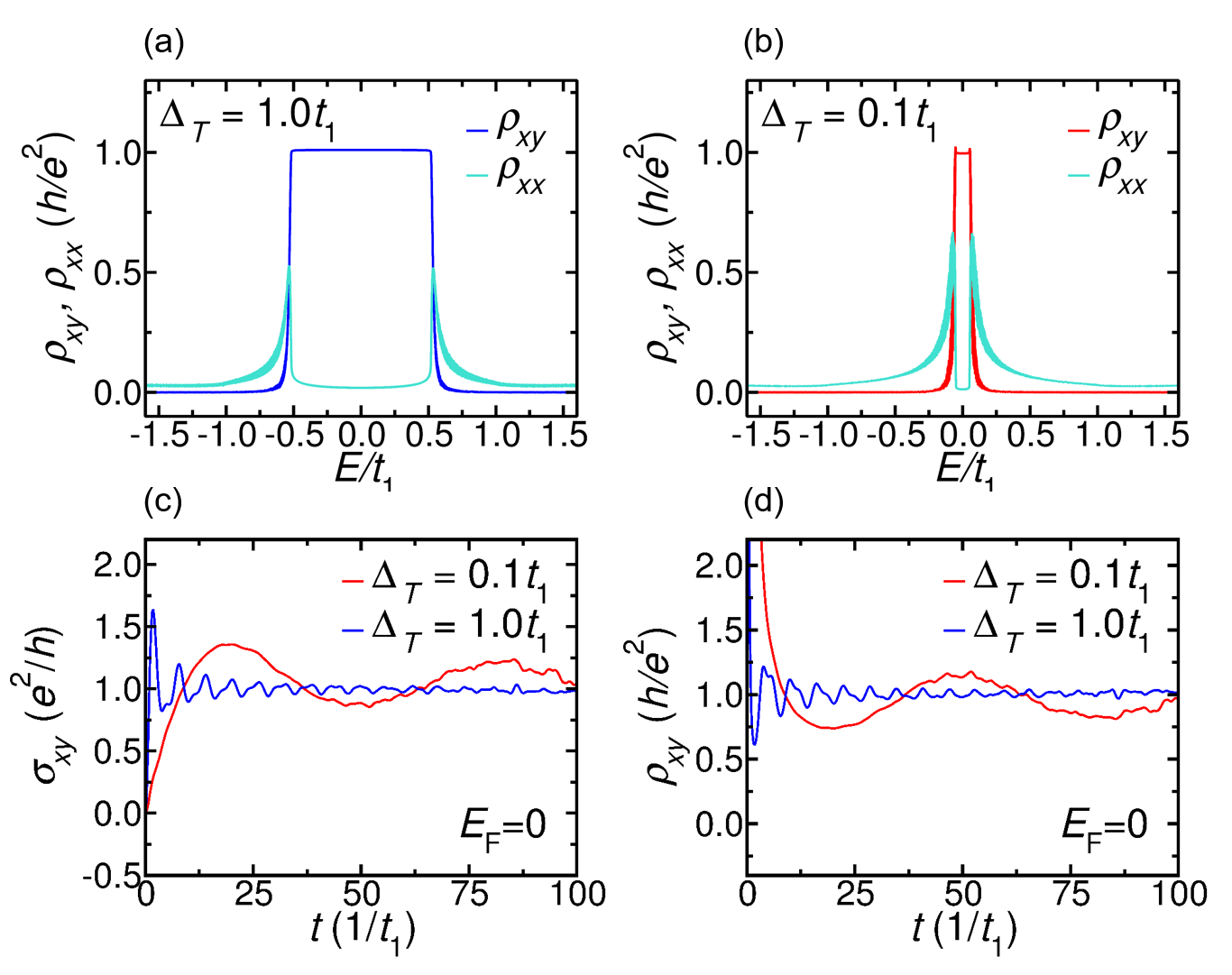}
\caption{(a) Longitudinal and transversal resistivity components of the Haldane model for graphene calculated from the displacement form of the Kubo conductivity. The model parameters are set to $\Delta_T=t_1$, $\Delta_{AB}=0$ and $V=0.1t_1$. (b) Same as in (a) but with reduced topological gap $\Delta_T=0.1t_1$. (c) and (d): Time-resolved Hall conductivity (c) and Hall resistivity (d) at the Dirac-point ($E=0$) for both topological gap sizes.}
\label{fig:Haldane_summary_v2}
\end{figure}

\paragraph{Numerical example: electrical conductivity tensor of the Haldane model.}
In this application, which is of foremost interest to us, the operators $\hat{A}$ and $\hat{B}$ are identified with the electric dipole moments along different Cartesian directions $e\hat{x}_{\alpha}$, while the modulation function is taken to be constant $F(t'-t)=F(0)$ and proportional to the electric field strength $E_{\beta}$. Then the linear response of the current density $\text{Tr}\left(\hat{\rho}(t)\hat{\jmath}_{\alpha}\right)=\sigma^{\text{dc}}_{\alpha\beta}E_{\beta}$ with $\hat{\jmath}_{\alpha}=e\dot{\hat{x}}_{\alpha}/V$ is described by the dc-conductivity tensor  
\begin{align}
\begin{aligned}
\sigma_{\alpha\beta}^{\text{dc}}&=\frac{e^2}{V}\lim\limits_{t\rightarrow\infty}\left(\frac{\beta}{4}\frac{\tan\left(\frac{\beta\hbar}{2}\frac{d}{dt}\right)}{\left(\frac{\beta\hbar}{2}\frac{d}{dt}\right)}\frac{d}{dt}\mathcal{D}^+_{x_\alpha x_\beta}(t)+\frac{1}{2\hbar}\mathcal{D}^-_{x_\alpha x_\beta}(t)\right)\,. \label{electrical conductiity tensor displacement form}
\end{aligned}
\end{align} 
We consider the classic Haldane model for honeycomb lattices \cite{Haldane1988} as a fundamental example of a topological Chern-insulator to demonstrate the theoretical and numerical description. Its Hamiltonian is defined on the graphene lattice as
\begin{align}
\begin{aligned}
\hat{H}_0&=-t_1\sum\limits_{\left<i,j\right>}\hat{c}^{\dagger}_i\hat{c}_j+t_2\sum\limits_{\left<\left<i,j\right>\right>}e^{i\phi_{ij}}\hat{c}^{\dagger}_i\hat{c}_j\\&+\frac{\Delta_{AB}}{2}\sum\limits_i\left(\delta_{iA}-\delta_{iB}\right)\hat{c}^{\dagger}_i\hat{c}_i+\sum\limits_iV_i\hat{c}^{\dagger}_i\hat{c}_i \label{Haldane Hamiltonian}
\end{aligned}
\end{align}
with the nearest neighbor coupling $t_1$  and the next nearest neighbor coupling $t_2$.  
The electronic onsite energies can be modified by an energy splitting $\Delta_{AB}$ that breaks the AB-sublattice symmetry, which however is taken $\Delta_{AB}=0$ for simplicity here. 
Additionally, a uniform Anderson disorder potential with $-V/2\leq V_i\leq V/2$ is applied. The next nearest neighbor coupling $t_2$ opens an energy gap $\Delta_T$ at the Dirac-point. This topological gap opens because time-reversal symmetry is broken leading to an anomalous quantum Hall effect. If the unimodular phase is set to $|\phi_{ij}|=\pi/2$   such that the total net flux inside a hexagon vanishes, the topological gap has a width of $\Delta_T=6\sqrt{3}t_2$. 
In Fig.~\ref{fig:Haldane_summary_v2} (a) and (b), we show the numerical results for the energy-resolved resistivities $\rho_{xy}$ and $\rho_{xx}$ for different values of $\Delta_T$, which are obtained from the time-resolved conductivity tensor in Eq.~\eqref{electrical conductiity tensor displacement form} taken at $t=20\pi/t_1$. The results inside the topological gap confirm the Hall conductivity $\sigma_{xy}=e^2/h$  and the Hall resistivity $\rho_{xy}=h/e^2$  in full consistency with the disorder-free case \cite{Haldane1988} and other implementations of the Kubo formula \cite{Garcia2015PRL} based on the Kubo-Bastin formula for the electrical conductivity \cite{BASTIN19711811}. 

In addition to the energy dependence of electron transport, we are here able to determine the time resolved dynamics of the conductivity and the resistivity, exemplarily shown in Fig.~\ref{fig:Haldane_summary_v2} (c) and (d). For example, the Hall plateau in $\sigma_{xy}$ and $\rho_{xy}$ emerges at early times and, for the large topological gap of $\Delta_T=t_1$, converges after 10 periods of $\tau_T=2\pi/\Delta_T$ with a root-mean-square deviation of 0.5\% relative to the analytical value (blue line in Fig.~\ref{fig:Haldane_summary_v2} (c) and (d)). This formation time would correspond to around 15 fs (if we take $t_1$=2.7 eV as a typical value for graphene). For the small gap of $\Delta_T=0.1t_1$, the formation process is tenfold slower and shows a stronger oscillatory behavior at early times (see red curve in Fig.~\ref{fig:Haldane_summary_v2} (c) and (d)). Still in both cases of significant Anderson disorder ($V=0.1t_1$), we conclude that the formation process of the topological state is stable against disorder-induced scattering. Slight differences in the convergence behaviour to a plateau value are observed between the gap center and the gap edges when the strength of the Anderson disorder is comparable to the system's topological gap (see Fig.~\ref{fig:Haldane_summary_v2} (b)).

We further emphasize that the transverse response is obtained at a very small fraction of the conventional simulation time because only $\mathcal{D}_{x_\alpha x_\beta}^+ (t)$ and $\mathcal{D}_{x_\alpha x_\beta}^- (t)$ need to be propagated numerically. 
As compared to other linear-scaling time-domain approaches \cite{OrtmannPRB2015,OrtmannRochePRL2011} that require a propagation of around 1,000--5,000 Lanczos vectors, the new time-domain approach demands the propagation of only 2--4 Lanczos vectors, which hence results in a speed-up and savings in computational time by 3 orders of magnitude.

Finally, the correspondence of the new formalism based on the expressions \eqref{Generalized Kubo displacement form} and \eqref{Generalized Einstein} to known forms of the Kubo formula for the electrical conductivity is compiled in the appendix for the interested reader. There, we also discuss further limiting cases, namely the high-temperature and the classical limit of Eqs.~\eqref{Generalized Kubo displacement form} and \eqref{Generalized Einstein}.

\paragraph{Conclusions.}
In this work, we have derived analytic forms of linear response functions by decomposition into time-symmetric and time-antisymmetric contributions. This enables an efficient implementation and computation of transverse linear response phenomena at the same footing as the longitudinal response. Different limiting cases that are known have been reproduced. This unified description makes the development of specific algorithms unnecessary and at the same time allows to study these responses in the time domain. As compared to other linear-scaling time-domain approaches to transverse responses, it benefits from a speed-up factor of 1000 or more. This allows the precise determination of topological effects in a time-domain approach that has not been established before.

\section{Acknowledgments}
We would like to thank the Deutsche Forschungsgemeinschaft for financial support (projects OR 349/1 and OR 349/3). Grants for computer time from the Zentrum f\"ur Informationsdienste und Hochleistungsrechnen
of TU Dresden (ZIH) and the Leibniz Supercomputing Centre in Garching
(SuperMUC-NG) are gratefully acknowledged.


\section{Appendix}
\subsection{Proof of Theorem 1 and Theorem 2}
As a first step to prove Theorem 1 and Theorem 2 in Eq.~\ref{Generalized Kubo displacement form} and  \eqref{Generalized Einstein}  we note that the response function $f_{AB}(t)$ can be expressed as
\begin{align}
f_{AB}(t)=\int\limits_0^{\beta}d\lambda\,g_{AB}(t+i\hbar\lambda)
\end{align}
with the complex function
\begin{align}
g_{AB}(t+i\hbar\lambda)=\text{Tr}\left(\hat{\rho}_0\dot{\hat{B}}(0)\hat{A}(t+i\hbar\lambda)\right) \label{g_AB definition}
\end{align}
and the inverse temperature $\beta=1/k_{\text{B}}T$ as the upper integration limit. The temperature enters the formalism via the equilibrium density operator $\hat{\rho}_0$, which corresponds to the canonical or the grand-canonical ensemble.
The dot in Eq.~\eqref{g_AB definition} denotes the time derivative with respect to $\hat{H}_0$, i.e.~$\dot{\hat{B}}(0)=(i/\hbar)\left[\hat{H}_0,\hat{B}(0)\right]$.
We use the intrinsic symmetries of
$f_{AB}(t)$ and $g_{AB}(t+i\hbar\lambda)$
\begin{align}
f_{AB}(t)&=-f_{BA}(-t)\,,\\
g_{AB}(t+i\hbar\lambda)^*&=-g_{BA}(-t+i\hbar\lambda)\,,
\end{align}
to show for the anti-symmetric part of the response function $f_{AB}^{\text{ta}}(t)=1/2\left(f_{AB}(t)-f_{AB}(-t)\right)$ the symmetry relation
\begin{align}
\begin{aligned}
f_{AB}^{\text{ta}}(t)&=\frac{1}{2}\left(f_{AB}(t)+f_{BA}(t)\right)\,,
\end{aligned}
\end{align} i.e. the function $f_{AB}^{\text{ta}}(t)$  is symmetric under the exchange of the operators $\hat{A}$ and $\hat{B}$. One can further show that
the time-antisymmetric part of the response function is given by the $\lambda$-integration of the real part of the complex function $g^{\text{ta}}_{AB}(t+i\hbar\lambda)$, i.e.
\begin{align}
\begin{aligned}
f_{AB}^{\text{ta}}(t)
&=\int\limits_0^{\beta}d\lambda \Re\left(g^{\text{ta}}_{AB}(t+i\hbar\lambda)\right)\,.
\label{f_as_integral_g}
\end{aligned}
\end{align}
Owing to $f_{AB}^{\text{ta}}(t)+f_{AB}^{\text{ta}}(-t)=0,$ we also have
\begin{align}
\begin{aligned}
0&=\int\limits_0^{\beta}d\lambda \Im\left(g^{\text{ta}}_{AB}(t+i\hbar\lambda)\right)\,.
\end{aligned}
\end{align}
From this we find that the $\lambda$-integration of $g_{AB}^{\text{ta}}(t+i\hbar\lambda)$ is equivalent to the integration over the real part only
\begin{align}
\int\limits_0^{\beta}d\lambda\,g_{AB}^{\text{ta}}(t+i\hbar\lambda)=\int\limits_0^{\beta}d\lambda\,\Re\left(g_{AB}^{\text{ta}}(t+i\hbar\lambda)\right)\,.
\end{align}

Now we rewrite $\Re\left(g_{AB}^{\text{ta}}(t+i\hbar\lambda)\right)$ by  applying the Baker-Campbell-Hausdorff formula to a time-dependent operator $\hat{O}(t+i\hbar\lambda)$
\begin{align}
\begin{aligned}
\hat{O}(t+i\hbar\lambda)&=e^{-\lambda\hat{H}_0}\hat{O}(t)e^{\lambda\hat{H}_0}\\&=\sum\limits_{k=0}^{\infty}\frac{\left(-\lambda\right)^k}{k!}\left[\hat{H}_0,\hat{O}(t)\right]_k
\end{aligned}
\end{align}
and substitute the nested commutators with the time derivative, i.e.~$\left[\hat{H}_0,\hat{O}(t)\right]_k=\left(-i\hbar\, d/dt\right)^k\hat{O}(t)$. We find
\begin{align}
\begin{aligned}
\Re\left(g_{AB}^{\text{ta}}(t+i\hbar\lambda)\right)
&=\cos\left(\hbar\lambda\frac{d}{dt}\right)\Re\left(g_{AB}^{\text{ta}}(t)\right)
\\&-\sin\left(\hbar\lambda\frac{d}{dt}\right)\Im\left(g_{AB}^{\text{ts}}(t)\right)\,.\label{Re(gAB(t+ihbarlambda)) in alternative form}
\end{aligned}
\end{align}   
We show next how the right hand side of \eqref{Re(gAB(t+ihbarlambda)) in alternative form} can be expressed with $f_{AB}^{\text{ta}}(t)$ and the displacement operator anticommutator function $\mathcal{D}^+_{AB}(t)$. We first find the equivalence
\begin{align}
\begin{aligned}  
\Re\left(g_{AB}^{\text{ta}}(t)\right)&=\frac{1}{4}\text{Tr}\left(\hat{\rho}_0\left(\dot{\hat{B}}(0)\hat{A}(t)+\dot{\hat{A}}(0)\hat{B}(t)\right.\right.\\&\left.\left.-\dot{\hat{B}}(0)\hat{A}(-t)-\dot{\hat{A}}(0)\hat{B}(-t)\right)\right)\\
&=\frac{1}{4}\frac{d}{dt}\mathcal{D}_{AB}^+(t)\,. \label{Re g from DAF}
\end{aligned}
\end{align}
For the imaginary part of $g_{AB}^{\text{ts}}(t)$ we obtain
\begin{align}
\begin{aligned}
i\Im\left(g_{AB}^{\text{ts}}(t)\right)&=\frac{1}{4}\text{Tr}\left(\hat{\rho}_0\left(\dot{\hat{B}}(0)\hat{A}(t)+\dot{\hat{A}}(0)\hat{B}(t)\right.\right.\\&\left.\left.+\dot{\hat{B}}(0)\hat{A}(-t)+\dot{\hat{A}}(0)\hat{B}(-t)\right)\right)\\
&=-\frac{i\hbar}{2}\frac{d}{dt}f_{AB}^{\text{ta}}(t)\,. 
\label{Im g from fsym}
\end{aligned}
\end{align}   
Inserting equations \eqref{Re g from DAF} and \eqref{Im g from fsym} into \eqref{Re(gAB(t+ihbarlambda)) in alternative form} one obtains    
\begin{align}
\begin{aligned}
\Re\left(g_{AB}^{\text{ta}}(t+i\hbar\lambda)\right)
&=\frac{1}{2}\frac{d}{d\lambda}\left(\frac{1}{2\hbar}\sin\left(\hbar\lambda\frac{d}{dt}\right)\mathcal{D}_{AB}^{+}(t)
\right.\\&\left.-\cos\left(\hbar\lambda\frac{d}{dt}\right)f_{AB}^{\text{ta}}(t)\right), \label{lambda derivative g_AB^sym}
\end{aligned}
\end{align}    
which, inserted into \eqref{f_as_integral_g}, allows integrating over $\lambda$. As a result, we obtain 
\begin{align}
\begin{aligned}
f_{AB}^{\text{ta}}(t)
&=\frac{1}{4\hbar}\sin\left(\beta\hbar\frac{d}{dt}\right)\mathcal{D}_{AB}^{+}(t)\\&
-\frac{1}{2}\left(\cos\left(\beta\hbar\frac{d}{dt}\right)-1\right)f_{AB}^{\text{ta}}(t),
\end{aligned}
\end{align}
which can be rewritten as   
\begin{align}
\frac{1}{2}\left(1+\cos\left(\beta\hbar\frac{d}{dt}\right)\right)f_{AB}^{\text{ta}}(t)=\frac{1}{4\hbar}\sin\left(\beta\hbar\frac{d}{dt}\right)\mathcal{D}_{AB}^{+}(t)\,.\label{solve f_AB 1}
\end{align}     
If we now bring the operator $1/2\left(1+\cos\left(\beta\hbar\,d/dt\right)\right)$ to the other side and apply the trigonometric identity $\sin(x)/(1+\cos(x))=\tan\left(x/2\right)$, we find the explicit representation of $f_{AB}^{\text{ta}}(t)$ in terms of the DAF $\mathcal{D}^+_{AB}(t)$   
\begin{align}
f_{AB}^{\text{ta}}(t)=\frac{1}{2\hbar}\tan\left(\frac{\beta\hbar}{2}\frac{d}{dt}\right)\mathcal{D}^+_{AB}(t)\label{f_AB in DAF form}
\end{align}
that holds for arbitrary linear response functions. 

For the time-symmetric part of the response function     
\begin{align}
\begin{aligned}
f^{\text{ts}}_{AB}(t)&=\frac{1}{2}\left(f_{AB}(t)+f_{AB}(-t)\right)
\label{time symmetric response function}
\end{aligned}
\end{align}   
we find   
\begin{align}
\begin{aligned}
f^{\text{ts}}_{AB}(t)&=-\frac{i}{2\hbar}\left(\text{Tr}\left(\hat{\rho}_0\left[\Delta\hat{A}(t),\Delta\hat{B}(t)\right]\right)\right.\\
&\left.+2\,\text{Tr}\left(\hat{\rho}_0\left[\hat{B}(0),\hat{A}(0)\right]\right)\right)\\
&=f_{AB}(0)+\frac{1}{2\hbar}\mathcal{D}^-_{AB}(t)\,.\label{time-symmetric response function displacement form 2}
\end{aligned}
\end{align}    
Thus, we have finally derived that the response function 
\begin{align}
\begin{aligned}
f_{AB}(t)&=f_{AB}^{\text{ts}}(t)+f_{AB}^{\text{ta}}(t)\\
&=f_{AB}(0)+\frac{1}{2\hbar}\mathcal{D}^-_{AB}(t)\\
&+\frac{1}{2\hbar}\tan\left(\frac{\beta\hbar}{2}\frac{d}{dt}\right)\mathcal{D}^+_{AB}(t)
\end{aligned}\label{correlation function with displacements}
\end{align}    
is solely expressed by the displacement operators and the constant offset $f_{AB}(0)$ at zero time (Theorem ~1). The general Kubo formula expressed in terms of the DAF and the DCF then reads
\begin{align}
\begin{aligned}
\text{Tr}\left(\hat{\rho}(t)\hat{A}\right)
&=\text{Tr}\left(\hat{\rho}_0\hat{A}\right)+\int\limits^{\infty}_0dt'\,F(t-t')\left(f_{AB}(0)\right.\\
&\left.+\frac{1}{2\hbar}\tan\left(\frac{\beta\hbar}{2}\frac{d}{dt'}\right)\mathcal{D}^+_{AB}(t')+\frac{1}{2\hbar}\mathcal{D}^-_{AB}(t')\right)
\label{general Kubo 3}\,.
\end{aligned}
\end{align}

In special cases when the operator of interest is of the form that it describes the time derivative with respect to $\hat{H}_0$, i.e.~$\dot{\hat{A}}=(i/\hbar)\left[\hat{H}_0,\hat{A}\right]$, we obtain for the initial condition that $\text{Tr}\left(\hat{\rho}_0\dot{\hat{A}}\right)=0$.
The response function can further be written as a time derivative 
\begin{align}
\begin{aligned}
f_{\dot{A}B}(t)&=-\frac{i}{\hbar}\text{Tr}\left(\hat{\rho}_0\left[\hat{B}(0),\frac{d}{dt}\hat{A}(t)\right]\right)\\
&=\frac{d}{dt}f_{AB}(t),
\end{aligned}
\end{align}
which is calculated based on Eq.~\eqref{Generalized Kubo displacement form}
\begin{align}
\begin{aligned}
f_{\dot{A}B}(t)&=\frac{1}{2\hbar}\tan\left(\frac{\beta\hbar}{2}\frac{d}{dt}\right)\frac{d}{dt}\mathcal{D}^+_{AB}(t)+\frac{1}{2\hbar}\frac{d}{dt}\mathcal{D}^-_{AB}(t)\,.
\end{aligned}
\end{align}
Inserting these findings into the linear response function for the operator $\dot{\hat{A}}$,
we arrive at the general expression of Eq.~\eqref{Generalized Einstein}
\begin{align}
\begin{aligned}
f_{\dot{A}B}(t)=\frac{1}{2\hbar}\frac{d}{dt}\mathcal{D}_{AB}^-(t)+\frac{1}{2\hbar}\tan\left(\frac{\beta\hbar}{2}\frac{d}{dt}\right)\frac{d}{dt}\mathcal{D}_{AB}^+(t),
\end{aligned}
\end{align}
which concludes the proof of Theorem 2.

\subsection{Proof of Theorem 3}
To prove the decomposition of the cross-correlation function $\mathcal{S}_{AB}(t)$ according to equations \eqref{Correlation time-symmetric} and  \eqref{Correlation time anti-symmetric} we start with the time-symmetric part $\mathcal{S}^{\text{ts}}_{AB}(t)$ that is defined as
\begin{align}
\begin{aligned}
\mathcal{S}_{AB}^{\text{ts}}(t)&=\frac{1}{2}\left(\mathcal{S}_{AB}(t)+\mathcal{S}_{AB}(-t)\right)\,.
\end{aligned}
\end{align}
The DAF $\mathcal{D}^{+}_{AB}(t)$ can be written as
\begin{align}
\begin{aligned}
\mathcal{D}^+_{AB}(t)
&=\text{Tr}\left(\hat{\rho}_0\left(\left\{\hat{A}(t),\hat{B}(t)\right\}-\left\{\hat{A}(t),\hat{B}(0)\right\}\right.\right.\\
&\left.\left.-\left\{\hat{A}(0),\hat{B}(t)\right\}+\left\{\hat{A}(0),\hat{B}(0)\right\}\right)\right)\\
&=2\text{Tr}\left(\hat{\rho}_0\left\{\hat{A}(0),\hat{B}(0)\right\}\right)-4\mathcal{S}_{AB}^{\text{ts}}(t)\\
&=4\left(\mathcal{S}_{AB}(0)-\mathcal{S}_{AB}^{\text{ts}}(t)\right)
\end{aligned}
\end{align}
that directly proves equation \eqref{Correlation time-symmetric}.
The proof of the time-antisymmetric part of the cross-correlation function $\mathcal{S}^{\text{ta}}_{AB}(t)$ turns out to be slightly more difficult. From the generic definition 
\begin{align}
\begin{aligned}
\mathcal{S}_{AB}^{\text{ta}}(t)&=\frac{1}{2}\left(\mathcal{S}_{AB}(t)-\mathcal{S}_{AB}(-t)\right)
\end{aligned}
\end{align}
we find
\begin{align}
\begin{aligned}
\mathcal{S}_{AB}^{\text{ta}}(t)&=\frac{1}{4}\text{Tr}\left(\hat{\rho}_0\left(\left\{\hat{B}(0),\hat{A}(t)\right\}-\left\{\hat{B}(t),\hat{A}(0)\right\}\right)\right)\,.
\end{aligned}
\end{align}

We derive the representation of the time-antisymmetric part of the cross-correlation function and use the following identity for the time-symmetric part of the response function $f_{AB}^{\text{ts}}(t)$
\begin{align}
f_{AB}^{\text{ts}}(t)=\int\limits_0 ^{\beta}d\lambda\,\Re\left(g_{AB}^{\text{ts}}(t+i\hbar\lambda)\right).
\end{align} 
Analogously to Eq.~\eqref{Re(gAB(t+ihbarlambda)) in alternative form}, we find for the integrand
\begin{align}
\begin{aligned}
\Re\left(g_{AB}^{\text{ts}}(t+i\hbar\lambda)\right)&=\cos\left(\hbar\lambda\frac{d}{dt}\right)\Re\left(g^{\text{ts}}_{AB}(t)\right)\\
&-\sin\left(\hbar\lambda\frac{d}{dt}\right)\Im\left(g^{\text{ta}}_{AB}(t)\right)\,, \label{Re(gAB(t+ihbarlambda))_ts in alternative form}
\end{aligned}
\end{align}
where the difference to Eq.~\eqref{Re(gAB(t+ihbarlambda)) in alternative form} is the interchange of the time-symmetric and time-antisymmetric parts.
We can further identify $\Re\left(g^{\text{ts}}_{AB}(t)\right)$ and $\Im\left(g^{\text{ta}}_{AB}(t)\right)$ as
\begin{align}
\Re\left(g^{\text{ts}}_{AB}(t)\right)&=-\frac{d}{dt}\mathcal{S}^{\text{ta}}_{AB}(t)\,,\\
i\Im\left(g^{\text{ta}}_{AB}(t)\right)&=-\frac{i\hbar}{2}\frac{d}{dt}f_{AB}^{\text{ts}}\,.
\end{align}
The further derivation follows steps that are analogous to the proof of Theorem 1 and results in the following relation between the time-antisymmetric part of the cross-correlation function $\mathcal{S}_{AB}^{\text{ta}}(t)$ and the time-symmetric part of the response function $f_{AB}^{\text{ts}}(t)$:
\begin{align}
f_{AB}^{\text{ts}}(t)=-\frac{2}{\hbar}\tan\left(\frac{\beta\hbar}{2}\frac{d}{dt}\right)\mathcal{S}_{AB}^{\text{ta}}(t)\,\label{time symmetric part or f_AB via time antisymmetric part of S_AB}.
\end{align}
Equating Eq.~\eqref{time-symmetric response function displacement form 2} and Eq.~\eqref{time symmetric part or f_AB via time antisymmetric part of S_AB} finally yields
\begin{align}
-\frac{2}{\hbar}\tan\left(\frac{\beta\hbar}{2}\frac{d}{dt}\right)\mathcal{S}_{AB}^{\text{ta}}(t)=f_{AB}(0)+\frac{1}{2\hbar}\mathcal{D}^-_{AB}(t)
\end{align}
and thus leads to Eq.~\eqref{Correlation time anti-symmetric} which proves Theorem 3.
Furthermore the fluctuation-dissipation theorem in the time domain is rediscovered as
\begin{align}
f_{AB}(t)=-\frac{2}{\hbar}\tan\left(\frac{\beta\hbar}{2}\frac{d}{dt}\right)\mathcal{S}_{AB}(t)
\end{align} 
by combining Theorem 1 and Theorem 3.

\subsection{Correspondence to known limits of the Kubo formula}
As an interesting side remark in this paper, we briefly show known limits of the conductivity, which can be obtained in absence of a magnetic field (in which $\mathcal{D}_{x_{\alpha}x_{\beta}}^-(t)=0$). By additionally considering the limit $\beta\hbar/2\, d/dt\rightarrow 0$, the electrical dc-conductivity tensor in Eq.~\eqref{electrical conductiity tensor displacement form} simplifies to the well-known form
\begin{align}
\sigma_{\alpha\beta}^{\text{dc}}=\frac{\beta e^2}{V}D_{\alpha\beta}
\end{align}
where we denote $D_{\alpha\beta}$ as the diffusivity tensor  $D_{\alpha\beta}=(1/4)\lim\limits_{t\rightarrow\infty}(d/dt)\mathcal{D}^+_{x_\alpha x_\beta}(t)$ in accordance with \cite{nakajima1958quantum}.
The so-defined tensor can be shown to be equivalent with its representation via the time-symmetric current-current correlation function  $\mathcal{S}^{\text{ts}}_{\jmath_{\alpha}\jmath_{\beta}}(t)=e^2\mathcal{S}^{\text{ts}}_{\dot{x}_{\alpha}\dot{x}_{\beta}}(t)$, which is the familiar expression \cite{Kubo1957,nakajima1958quantum}
\begin{align}
\frac{\beta e^2}{4V}\lim\limits_{t\rightarrow\infty}\frac{d}{dt}\mathcal{D}^+_{x_{\alpha}x_{\beta}}(t)=\frac{\beta }{V}\lim\limits_{t\rightarrow\infty}\int\limits_0^tdt'\mathcal{S}^{\text{ts}}_{\jmath_\alpha \jmath_\beta}(t')\,.
\end{align}
This conductivity tensor is symmetric in the displacement operators and may possesses off-diagonal tensor components even in absence of magnetic fields, which are associated with off-diagonal diffusion coefficients. The diagonal components of the tensor simply read
\begin{align}
\begin{aligned}
\sigma^{\text{dc}}_{\alpha\alpha}&=\frac{\beta e^2}{2V}\lim\limits_{t\rightarrow\infty}\frac{d}{dt}\text{Tr}\left(\hat{\rho}_0\left(\Delta\hat{x}_{\alpha}(t)\right)^2\right)\\
&=\frac{\beta}{2V}\int\limits_{-\infty}^{\infty}dt\text{Tr}\left(\hat{\rho}_0\hat{\jmath}_{\alpha}(0)\hat{\jmath}_{\alpha}(t)\right),
\end{aligned}
\end{align}
which represent two well-known Kubo formulae for the longitudinal dc-conductivity based on the average mean square displacements or the current-current correlation function, respectively.

\subsection{High-temperature and classical limit of the response function.}
In this work, we identify the high-temperature limit of the response function $f_{AB}(t)$ and $f_{\dot{A}B}(t)$ in Eqs.~\eqref{Generalized Kubo displacement form} and \eqref{Generalized Einstein} by letting $\beta\hbar/2 \,d/dt\rightarrow 0$, which yields  $
f^{\text{high-}T}_{AB}(t)=(\beta/4)\left((d/dt)\mathcal{D}^+_{AB}(t)+\mathcal{D}^+_{\dot{A}B}(t)-4\mathcal{S}_{\dot{A}B}(0)\right)$ and $
f^{\text{high-}T}_{\dot{A}B}(t)=(\beta/4\,d/dt)\left((d/dt)\mathcal{D}^+_{AB}(t)+\mathcal{D}^+_{\dot{A}B}(t)\right)
$, respectively. This describes the limit where the energy differences between quantum states become small against the temperature. Interestingly, $f_{AB}(t)$  can be expressed as a simple combination of DAFs without using the DCF, which is in line with the transition to classical statistical physics where commutators are neglected. However, it should not be confused with the classical limit of $\hbar\rightarrow 0$ since the DAFs account for the full quantum mechanical description including all orders of quantum interference processes. 

To obtain the classical limit of the above response function, we substitute the operators $\hat{A}(t)\rightarrow A(t)$, $\hat{B}(t)\rightarrow B(t)$, and $\hat{\rho}_0\rightarrow \rho_0$ with their classical observables defined on phase space. According to this quantum-classical correspondence, the trace over the quantum statistical ensemble is replaced with the continuous integration over the phase space $\text{Tr}(\hat{\rho}_0 (...))\rightarrow(1/2\pi\hbar)^{3N}\int d^{3N} pd^{3N} x\, \rho_0({\bf{x}},{\bf{p}} )(...)=\left<\rho_0...\right>$.
Then we find for the classical limit
\begin{align}f_{AB}^{\text{class}}(t)&=\frac{\beta}{4}\left(\frac{d}{dt}\mathcal{D}_{AB}^{+,\text{class}} (t)+\mathcal{D}_{\dot{A}B}^{+,\text{class}}(t)-4\mathcal{S}_{\dot{A}B}^\text{class}(0)\right)\,,\\
f_{\dot{A}B}^{\text{class}}(t)&=\frac{\beta}{4}\frac{d}{dt}\left(\frac{d}{dt}\mathcal{D}_{AB}^{+,\text{class}} (t)+\mathcal{D}_{\dot{A}B}^{+,\text{class}}(t)\right)
\end{align}
with the classical version of the corresponding DAFs 
\begin{align}
\mathcal{D}_{AB}^{+,\text{class}}(t)&=2\left<\rho_0\Delta A(t)\Delta B(t)\right>\,,\\
\mathcal{D}_{\dot{A}B}^{+,\text{class}}(t)&=2\left<\rho_0\Delta \dot{A}(t)\Delta B(t)\right>
\end{align}
where  $\Delta A(t)=\Delta A({\bf{x}} (t),{\bf{p}} (t))$ denotes the classical displacement of the observable $A$ along the trajectories ${\bf{x}}(t)$ and ${\bf{p}}(t)$ that solve the classical equations of motion according to the Hamilton function $H_0 ({\bf{x}},\bf{p})$ of the ensemble. The classical displacement of the observable $\dot{A}$ is defined as $\Delta\dot{A}(t)=\{H_0,\Delta A(t)\}$ where $\{...\}$ denotes the Poisson bracket.
\end{document}